\documentclass[aps,pre,preprint,groupedaddress]{revtex4}

\usepackage{graphics}

\begin{document}


\title{A Group-Based Yule Model for Bipartite Author-Paper Networks}

\author{Michel L. Goldstein}
\email[]{mgoldst@okstate.edu}
\author{Steven A. Morris}
\email[]{samorri@okstate.edu}
\author{Gary G. Yen}
\email[]{gyen@okstate.edu}
\affiliation{
Oklahoma State University\\
Electrical and Computer Engineering\\
Stillwater, OK 74078, USA }

\date{\today}

\begin{abstract}
This paper presents a novel model for author-paper networks, which
is based on the assumption that authors are organized into groups
and that, for each research topic, the number of papers published by
a group is based on a success-breeds-success model. Collaboration
between groups is modeled as random invitations from a group to an
outside member. To analyze the model, a number of different metrics
that can be obtained in author-paper networks were extracted. A
simulation example shows that this model can effectively mimic the
behavior of a real-world author-paper network, extracted from a
collection of 900 journal papers in the field of complex networks.
\end{abstract}


\maketitle

\section{Introduction}

This paper presents a realistic bipartite model of author-paper
networks, a phenomenon which has been studied since the 1920s
\cite{lotka26frequency}.  The proposed growth model is based on
modeling groups of authors using a `nested' Yule process
\cite{yule24mathematical}, and further models `loose ties' among
author groups as a Watts-Strogatz small world process
\cite{watts98collective}. The full bipartite representation of the
network allows construction of many meaningful metrics to evaluate
the validity of the proposed model against actual author-paper
networks. Using a collection of 900 papers covering the topic of
complex networks, we will show that the proposed model faithfully
reproduces the characteristics of six metrics: 1) authors per paper
distribution, 2) papers per author distribution (Lotka's Law), 3)
co-author clustering coefficient distribution, 4) co-authorship per
author pair distribution, 5) collaborator per author distribution,
and 6) minimum path between author pairs distribution.

The model and the validation metrics presented in this paper are
innovative when considered against previous models of Lotka's Law or
models of author collaboration networks. Lotka's Law models deal
with single authors without modeling collaboration, while
collaboration models cannot describe Lotka's Law and single authors.
Both types of models are usually validated against simple power law
link degree distributions: papers per author for testing Lotka's Law
models, or collaborators per author for testing author collaboration
models. Power law link degree distributions are easy to duplicate
using several types of processes \cite{albert02statistical}. Because
of this, such simple models offer little insight into underlying
processes that generate author-paper networks.

The proposed model, which deals with groups of authors rather than
single authors, reveals the importance of research workgroups
(author groups) in author-paper networks. The model indicates that
publication by author groups is driven by a success-breeds-success
(SBS) process, and further, that authorship by single authors within
these groups is a SBS process as well. Yet, surprisingly, intergroup
collaboration, i.e. loose ties, appears to be well modeled by a
small world network of random interlinkages.

Bipartite author-paper networks are formed by two types of entities,
the authors and papers, and the authorship links between them. There
exists much analysis in the literature on the features of real-world
author-paper networks. The first of these analyses were presented by
Lotka \cite{lotka26frequency}. His analysis, which contained a
dataset of journal articles compiled by hand, showed that the
distribution of the number of papers per author follows a zeta
distribution, a pure power-law, with an exponent of approximately 2.
This observation is currently referred to as the Lotka's Law of
Scientific Productivity. A large number of other studies reinforced
the power-law concept for the number of papers per author
distribution, especially when considering only the tail of the
distribution. These studies show that the observed exponent varies
with the dataset \cite{pao85lotka,newman00who}.

The observation of this distribution is very important, but it does
not explicitly provide an insight into network dynamics. For this, a
dynamic growth model is needed. Of the dynamic models in the
literature, almost all are evaluated using crude comparisons to
simple paper per author distributions and ignore other important
metrics, such as clustering coefficient distribution or collaborator
distribution. A complete and useful model must be able to mimic the
real behavior of the author-paper network across many important
network metrics.

This paper provides a new model for the growth of author-paper
networks and a step-by-step presentation of the important features
of a real-world author-paper network that a model has to mimic. The
proposed model, although very simple, approximates well all these
features, thus building confidence in the validity of the model and
the insight that the model provides into the actual dynamics of
real-world author-paper networks.

\section{Author-Paper Network Models}

A number of different bipartite author-paper models exist in the
literature. These models attempt to explain the process generating
the power-law distribution of the papers per author distribution.
They are fundamentally different from the usual preferential
connection models, such as the Barab\'asi-Albert model
\cite{albert02statistical}, because they model bipartite networks,
in which one partition contains all authors and the other all
papers. Although it is possible to transform a bipartite network
into a simple graph by projection \cite{dorogovtsev02evolution},
this transformation removes the ability to calculate metrics to
evaluate the validity of the model.

In the model presented by Newman \emph{et al.}
\cite{newman01random}, the goal was to enable the generation of any
degree distributions, such as Poisson, exponential and power-law,
for simple, directed and bipartite graphs. The proposed method is
very general but it is mainly focused on predicting three features:
the average degree, the clustering coefficient
\cite{watts98collective}, and the degree distribution of the
projected graph. The model is able to effectively predict the
features for a network of company directors, however it fails to
approximate the features of authorship networks.

Huber \cite{huber02new} presents a model of authors to predict five
different features: the rate of production, career duration,
randomness, Poisson-ness distribution (related to the variance of
the author's productivity through time) and the distribution of
papers per author. Huber's model is complex and involves
distributions of career durations (assumed exponential) and Poisson
distributed counts of papers, based on the author's productivity.
Although this model predicts very well the features of interest, its
major drawback is that it does not model the existence of
co-authors. In the model, each author is ``evolved'' individually. A
useful model must have the ability to predict collaboration
patterns.

Recently, B\"orner \emph{et al.} \cite{borner04simultaneous}
presented a model in which the author network and the reference
networks evolve simultaneously. This study is an important
acknowledgement that multiple interconnected networks exist in
collections of journal papers, and that the challenge of modeling
such paper collections is to find the basic rules of author behavior
that produce the growth characteristics of the multiple
interconnected networks contained in them.  B\"orner \emph{et al}'s
main goal was to predict the evolution of the number of papers,
authors and citations in a large and heterogeneous collection of
journal articles, such as all of the papers published in the
Proceedings of the National Academy of Science from 1981 to 2001.
The paper includes a detailed set of proposed author behavioral
rules and predicts gross measures of author, paper, and reference
growth well, but the study does not discuss detailed metrics of
network characterization.

One major disadvantage of all models found in the literature is the
inability to predict most of the features of real-world networks.
The prediction of only one or two features greatly weakens the
usefulness of such models as models of real-world behavior.

\section{Proposed Group-Based Yule Model}

A Yule model is a preferential connection process first proposed as
a model of biological evolution by Yule in 1924
\cite{yule24mathematical}. Our model uses a Yule process to model
the growth of author groups in the author-paper network. The
proposed model is based on the observation that usually authors are
part of a research group. Most of the papers they write are
co-authored with other members of their group. Collaboration between
research groups happens, but multi-group papers are far less common
than in-group papers.

\begin{figure}
\resizebox{0.45\textwidth}{!}{%
\includegraphics{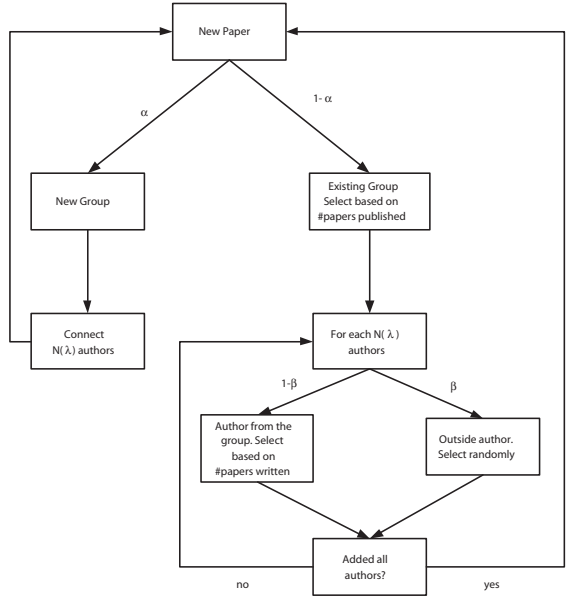}}%
\caption{Diagram of the proposed group-based Yule model for
author-paper networks\label{model}}
\end{figure}

A diagram of the model can be seen in Fig.\ \ref{model}. When a
paper is created there is a probability $\alpha$ that a new author
group is created with $N_g$ all new members, where $N_g$ is a
constant. The number of authors of the paper, $N(\lambda)$, is the
first author plus a Poisson-distributed number of additional
authors. This 1-shifted-Poisson distribution has parameter
$\lambda$. The probability distribution of the 1-shifted-Poisson,
$p_{sp}(k)$, is given in Eq.\ \ref{eq:spoisson}.

\begin{equation}
p_{sp}(k)=\frac{\lambda^{(k-1)}e^{-\lambda}}{(k-1)!},\ \ \ \ \
  k=\{1,2,\ldots\}, \label{eq:spoisson}\end{equation} where $k$ is the
number of authors and $p_{sp}(k)$ is the probability of a paper
having $k$ authors.

If a new group is not created, an existing author group is chosen
using the following probability distribution:

\begin{equation}
p_g(q) = \frac{q}{N_p},\end{equation} where $q$ is the number of
papers that this group has published, $N_p$ is the total number of
papers in the network, and $p_g(q)$ is the probability of an
existing group authoring a paper. This is the Yule process which
favors groups in proportion to the number of papers they have
published.

When an existing group is selected, it is necessary to select the
authors within the group that author the paper. The number of
authors of the paper is modeled as a 1-shifted Poisson distribution.
In order to model interconnection between groups (`loose ties'), for
each author, there is a probability $\beta$ that this author is from
another group. If so, the author is selected randomly from among all
authors in the network, whether they have authored a paper or not.
If an outside author is not chosen, an author from the selected
group is chosen. This selection is done by another preferential
connection process, modified to allow selection of authors that have
never published a paper. The probability of selecting an author $i$
in the group is:

\begin{equation}
p_a(i) = \frac{k_i+1}{\sum{k_j}+N_g}, \end{equation} where $k_i$ is
the number of papers written by author $i$, $\sum{k_j}$ is the sum
of the number of authorships among the authors in the group and
$N_g$ is the number of authors in the group. This is a preferential
attachment process which favors authors by the number of papers they
have previously published.

The paper creation cycle of Fig.\ \ref{model} repeats until the
desired number of papers is added to the network.

In summary, this model has four parameters: the group size, $N_g$,
assumed always constant for this simple model; the probability of
creating a new group, $\alpha$; the probability of choosing an
author from another group, $\beta$; and the Poisson parameter that
defines the distribution of number of authors per paper $\lambda$.
Given a dataset to be modeled, it is easy to analytically determine
$\alpha$ and $\lambda$.

The following section presents methods for obtaining these
parameters to model a real-world network. Methods for correctly
validating the model are also presented, by analyzing network
metrics.

\section{Example:}

The example is a collection of papers covering the specialty of
complex networks. This data set, collected from the Science Citation
Index, contains 900 papers, 1,354 authors, and 2,274 authorships
linking authors to papers. The first parameter, $\alpha$, is
obtained by determining the probability of new group creation. This
probability is estimated using a paper-by-paper pass through the
network to determine the fraction of papers that appeared with a
completely new set of authors.

The parameter $\lambda$ is calculated by dividing the total number
of authorships by the number of papers and subtracting 1
(1-shifted-Poisson). The number of authors per group, $N_g$, was
chosen heuristically as 20, which is assumed as the upper limit of
the number of researchers that can efficiently interact as a group.

The `loose tie' parameter $\beta$ is estimated by matching the
co-authorship distribution (the distribution of the number of times
pairs of authors have co-authored) by trial-and-error. The matching
of the co-authorship distribution will be explained below.

The parameters estimated for the example network are:

\begin{itemize}
    \item $\hat{\alpha}=0.33$
    \item $\hat{\beta}=0.1$
    \item $\hat{\lambda}=1.527$
    \item $N_G=20$
\end{itemize}

\begin{figure}
\resizebox{0.45\textwidth}{!}{%
\includegraphics{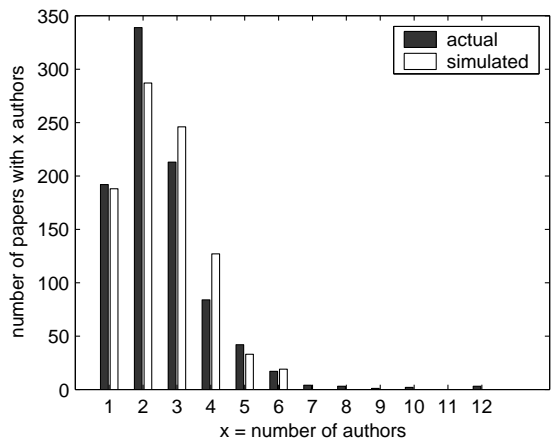}}%
\caption{Frequency distribution comparison for the number of authors
per paper between the actual distribution and the simulated
distribution. $\lambda_{actual}=1.527$,
$\lambda_{sim}=1.651$.\label{fig:ap}}
\end{figure}

\begin{figure}
\resizebox{0.45\textwidth}{!}{%
\includegraphics{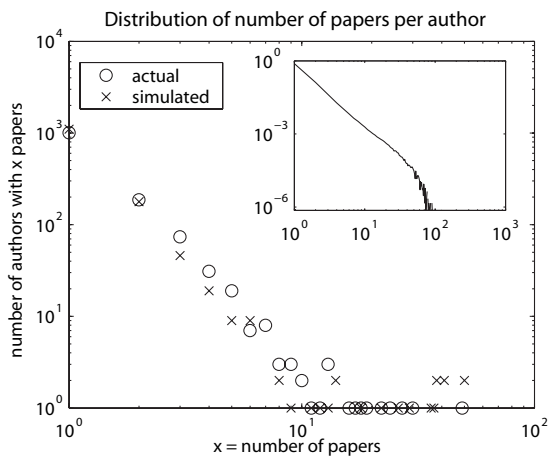}}%
\caption{Frequency distribution comparison for the number of papers
per author, Lotka's Law, between the actual distribution and the
simulated distribution. The actual distribution has a power-law
exponent of $\gamma=2.544$ and the simulated distribution has
$\gamma=2.770$. The inset shows the model-predicted paper per author
distribution, which fits a zeta distribution.\label{fig:pa}}
\end{figure}

\begin{figure}
\resizebox{0.45\textwidth}{!}{%
\includegraphics{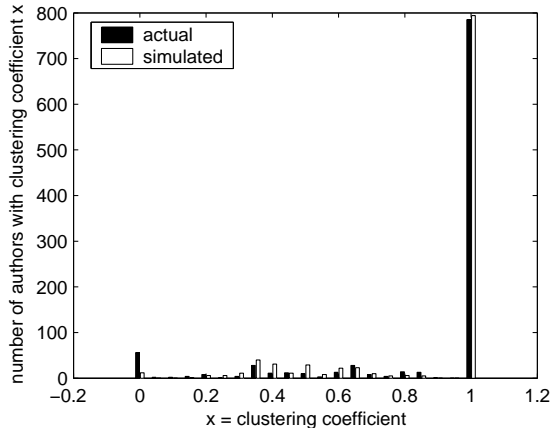}}%
\caption{Frequency distribution comparison of the clustering
coefficient. $C_{actual}=0.867$, $C_{sim}=0.881$.\label{fig:cc}}
\end{figure}

\begin{figure}
\resizebox{0.45\textwidth}{!}{%
\includegraphics{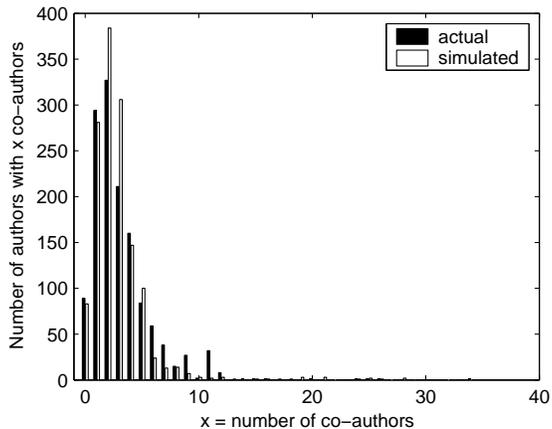}}%
\caption{Frequency distribution comparison of the number of
collaborators per author. $\mu_{actual}=3.15$, $\mu_{sim}=2.82$.
\label{fig:colab}}
\end{figure}

\begin{figure}
\resizebox{0.45\textwidth}{!}{%
\includegraphics{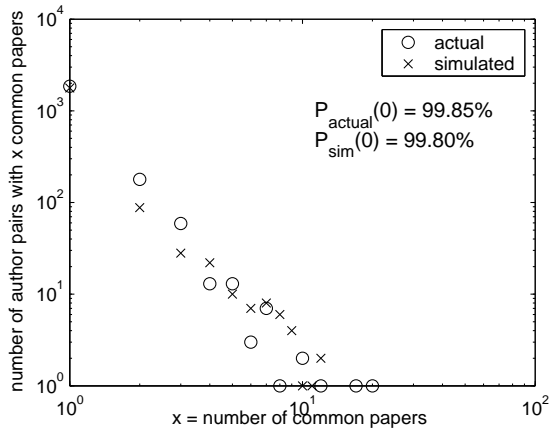}}%
\caption{Frequency distribution comparison of the co-authorship
distribution, showing the number of papers co-authored by each pair
of authors.\label{fig:coa}}
\end{figure}

\begin{figure}
\resizebox{0.45\textwidth}{!}{%
\includegraphics{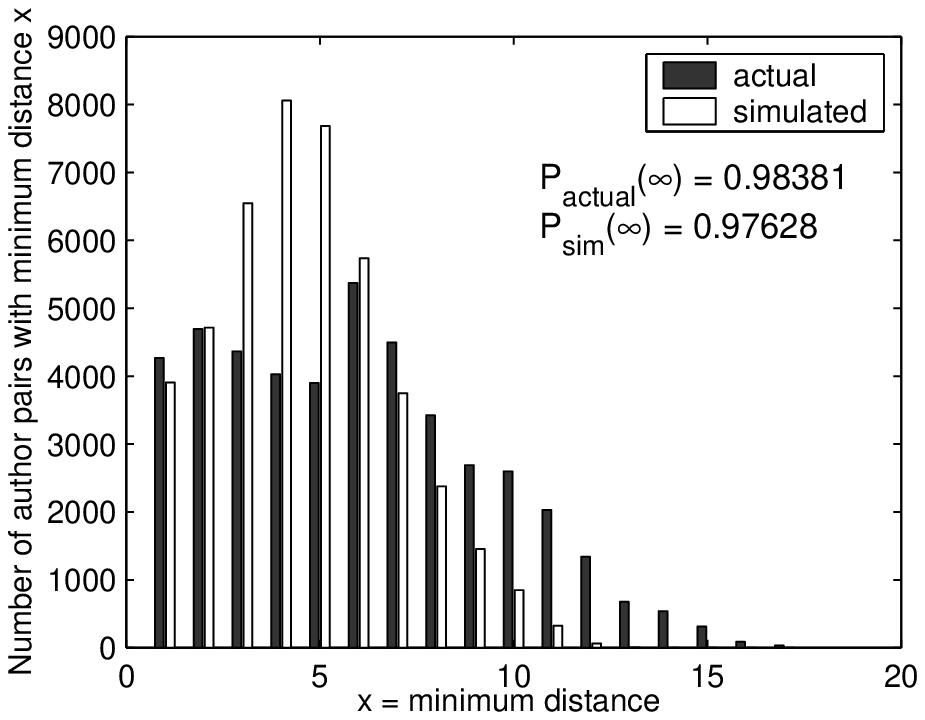}}%
\caption{Frequency distribution comparison of the minimum distance
between authors, i.e. the minimum number of links between each pair
of authors.\label{fig:md}}
\end{figure}

To validate the model, several metrics are used to compare model
simulations to the actual network. The following metrics are used
for comparison:

\begin{description}
    \item[Authors per paper] The distribution of the number of
    authors per paper. As discussed above, this is simulated as 1-shifted Poisson
    distributed. Note in Fig.\ \ref{fig:ap} the close match of
    actual to simulated distributions. This metric is important
    because it predicts the mean number of participants on projects
    within the group, an important measure of interaction within
    workgroups.
    \item[Papers per author distribution (Lotka's Law)] This is the
    distribution of the number of papers that each author published.
    Note in Fig.\ \ref{fig:pa} the close match of simulated
    frequencies to actual frequencies for this metric. This metric
    is important because it measures the distribution of
    productivity among authors in a specialty, modeling the
    formation of core groups of researchers in a specialty. The
    inset in Fig.\ \ref{fig:pa} shows the model-predicted paper per
    author distribution, generated by gathering statistics from
    1,000 simulations. The predicted distribution certainly models
    Lotka's Law, producing an excellent fit to a zeta distribution
    with an exponent of 2.77. Fitting was done using Maximum
    Likelihood Expectation and the fit passed a Kolmogorov-Smirnov
    (KS) test \cite{goldstein04problems} at an Observed Significance
    Level (OSL) of $10\% < OSL < 1\%$, $T=0.0031$, $N=1.3 \times 10^6$
    .
    \item[Co-author clustering coefficient distribution] The clustering coefficient was
    first introduced by Watts and Strogatz \cite{watts98collective}
    as a scalar mean clustering coefficient. However, when observing
    the distribution of the clustering coefficients, a very
    interesting characteristic is found in co-author networks: a
    large spike at unity. Therefore, it is imperative to use the
    distribution as the metric rather than the mean, which effectively
    hides unity spike behavior. For example, although author
    networks usually have a mean clustering coefficient of 0.8,
    comparable to that of citation networks \cite{albert02statistical}, the
    distribution of the co-author networks clustering coefficient is fundamentally different from the
    distribution of clustering coefficient in citation networks \cite{morris04manifestation}.
    Newman discusses this distribution in \cite{newman01random} and
    models it, with limited success. Note in Fig.\
    \ref{fig:cc} that simulation using the proposed model
    fully mimics the distribution of the clustering
    coefficient. This metric is important because it measures the
    tendency of authors to work in local groups.
    \item[Collaborator distribution] The distribution of the number
    of unique co-authors to each author in the network. Newman
    attempted to model this distribution with only partial success
    \cite{newman01random}. Note the close match of the simulated to
    actual co-authorship frequencies in Fig.\ \ref{fig:colab}. This
    metric is important because it measures the tendency of authors
    to work with other authors.
    \item[Co-authorship distribution] This is the distribution of the number of
    common papers between pairs of authors, across all author pairs in the network.
    Fig.\ \ref{fig:coa} shows that the proposed model matches the actual
    distribution well. This is an important metric because it
    measures the tendency of pairs of authors to repeatedly work together on
    individual projects.
    \item[Minimum distance distribution] Fig.\ \ref{fig:md} shows the distribution of the
    minimum distance between a pair of authors in the network, i.e.
    the minimum length of the path of co-authorships between them.
    This metric is important because it measures the tendency of groups to invite outside
    workers on projects.
\end{description}

For additional discussions of network metrics applicable to
author-paper networks, see Newman \cite{newman00who}, who discusses
several of the metrics used here.

All metrics shown above present a close match between the real-world
network and the model simulation. As an exception, the minimum path
distribution shows a fair amount of deviation, but this distribution
appears to be unstable and tends to change greatly from simulation
to simulation. The actual minimum path length distribution is
probably unstable as well, but investigation of that hypothesis is
outside the scope of this paper.

\section{Conclusions}

This paper proposes a very simple model for author-paper networks by
introducing the concept of preferential attachment of group
authoring of papers. Adding this simple concept to a Yule-type
process it was possible to obtain very similar behavior using
multiple metrics, when comparing to a real-world network. This
suggests that, in the real world, the modeling of research groups is
essential to understand the dynamics of paper authoring. Analysis of
single authors or random connections between authors, as proposed by
previous researchers, do not provide a reasonable model of reality.

Another important conclusion drawn from this model is that `loose
ties' between groups is well modeled by simple random inter-group
co-authorships. This implies that group collaboration does not
actually work by establishing formal long-term commitments, but by
single collaborations, possibly from informal meetings at
conferences, or e-mail discussion lists. Multiple collaboration with
outside groups may happen in real life, but such collaborations are
uncommon and do not affect the gross characteristics of the network.
This model further implies that outside collaboration is not
dependent on the amount of work that the outside person has done in
the field.

Note that while there is local preferential connection of authors
within groups, and global preferential connection of the groups
themselves, the inter-group linkage approximates a Watts-Strogatz
small world process. The model here is really a hybrid, being a
``nested preferential connection, global small world'' model.

We also showed that using only a single metric, such as the
distribution of papers per author, or a single mean value for the
clustering coefficient, incompletely validates a model. Analyzing
multiple metrics, allows validation against specific behaviors that
fully characterize the network.

It is important to note that this model only accounts for the
behavior of authorships in a collection of papers. To actually
understand the nature of collections of journal papers it would be
necessary to implement and discuss the interaction of this
author-paper bipartite network with the other bipartite networks in
the paper collection, such as the paper-reference network
\cite{naranan71power,redner98how,morris04manifestation},
paper-journal network (Bradford's Law) \cite{white98bibliometrics},
and paper-term network (Zipf's Law) \cite{white98bibliometrics}. The
analysis of their complex interaction will certainly shed light on a
large number of open questions regarding the growth and mapping of
information structures.

\bibliography{authorship}

\end{document}